\documentclass{article}
\usepackage{emulateapj}

\begin{document}

\submitted{Submitted to ApJ Letters, Oct. 11, 1999}

\title{Compton Dragged Gamma--Ray Bursts associated with Supernovae}

\author{Davide Lazzati\altaffilmark{1}, Gabriele Ghisellini}

\affil{Osservatorio Astronomico di Brera, Via Bianchi 46, I--23807
Merate (Lc), Italy}

\altaffiltext{1}{Dipartimento di Fisica, Universit\`a degli Studi di Milano,
Via Celoria 16, I--20133 Milano, Italy}

\author{Annalisa Celotti}

\affil{Scuola Internazionale Superiore di Studi Avanzati, via Beirut 2/4, 
I--34014 Trieste, Italy}

\and \author{Martin J. Rees}

\affil{Institute of Astronomy, University of Cambridge, Madingley Road, 
Cambridge CB3 0HA, U.K.}

\begin{abstract}
It is proposed that the gamma--ray photons that characterize the prompt 
emission of Gamma--Ray Bursts are produced through the Compton drag
process, caused by the interaction of a relativistic fireball with
a very dense soft photon bath. If gamma--ray bursts are indeed 
associated with Supernovae, then the exploding star can provide enough 
soft photons for radiative drag to be effective.
This model accounts for the basic properties of gamma--ray bursts, i.e.
the overall energetics, the peak frequency of the spectrum
and the fast variability, with an efficiency which can exceed 50\%.
In this scenario there is no need for particle acceleration in relativistic 
collisionless shocks. 
Furthermore, though Poynting flux may be important in accelerating
the outflow, no magnetic field is required in the gamma--ray production.
The drag also naturally limits the relativistic expansion of 
the fireball to $\Gamma \lesssim 10^4$.
\end{abstract}

\keywords{
gamma rays: bursts ---
supernovae: general ---
radiation mechanisms: non--thermal
}

\section{Introduction}
%

In the leading scenario for GRBs and afterglows, the gamma--ray event
is produced by internal shocks in a hyper--relativistic inhomogeneous
wind (Rees \& M\'esz\'aros 1994) while the afterglow
is produced as the fireball drives a shock wave in the external
interstellar medium (M\'esz\'aros \& Rees 1997).  Even if there is a
large consensus that both gamma--rays and afterglow
photons are produced by the synchrotron process, recently some doubts
have been cast on the synchrotron interpretation for the burst itself
(Liang 1997; Ghisellini \& Celotti 1998; Ghisellini, Celotti \& Lazzati 1999).

The nature of the progenitor is still a matter of active debate, 
as the sudden release of 
a huge amount of energy in a compact region generating a fireball
does not keep trace of the way this energy has been produced. 
For this reason, the study of the interactions of the fireball with the 
surrounding medium seems to be the most powerful mean to unveil 
the GRB progenitor.
At least two models are in competition: the merging of a binary system 
composed of two compact objects (Eichler et al. 1989) and the 
Hypernova--Collapsar model (Woosley 1993, Paczy\'nski 1998), i.e. 
the core collapse of a very massive star to form a black hole.

After the discovery and the multiwavelength observations of many afterglows,
circumstantial evidence has accumulated for GRB exploding in dense
regions, associated to supernova--like phenomena.
In fact, (a) host galaxies have been detected in many cases 
(Sahu et al. 1997; see Wheeler 1999 for a review),
and some of them show starburst activity (Djorgovski et al. 1998, 
Hogg \& Fructher 1999);
(b) large hydrogen column densities have sometimes been detected in X--ray 
afterglows (Owens et al. 1998); 
(c) non--detections of several X--ray afterglows in the optical band can be 
due to dust absorption (Paczy\'nski 1998); (d) a possible iron line 
feature has been detected in the X--ray afterglow
of GRB~970508 (Piro et al. 1999, Lazzati et al. 1999) and 
(e) the rapid decay with time of several afterglows can be explained by
the presence of a pre--explosion wind from a very massive star
(Chevalier \& Li 1999).
More recently, the possible presence of supernova (SN) emission in 
the late afterglows light curves
of GRB~970228 (Galama et al. 1999, Reichert 1999)
and GRB~980326 (Bloom et al. 1999) has added support in favor of the
association of some GRBs with the final evolutionary stages of massive stars.

Although in these models the available energy is larger than in the case 
of compact binary mergers, the very small efficiency of internal shocks 
(see, e.g., Spada, Panaitescu \& M\'esz\'aros 1999)
seems to be inconsistent with the fact that more energy 
can be released during the burst proper than the afterglow 
(Paczy\'nski 1999, see also Kumar \& Piran 1999).

In this letter we show that if GRBs are associated with supernovae, 
Compton drag inside the relativistic wind can produce 
both the expected energetics and the peak energy 
of the spectrum of a classical long duration GRB. 
In this new scenario the efficiency is not limited by internal 
shock interactions, and the successful modeling of afterglows 
with external shocks is left unaffected.

The Compton drag effect has been already invoked for GRBs by Zdziarski et 
al. (1991) and Shemi (1994). Cosmic background radiation (at high redshift), 
central regions of globular clusters and AGNs were identified as plausible
sources of soft photons, but none of these scenarii was able to
account for all the main properties of GRBs. However, the growing evidence 
of association of GRB explosions with star--forming regions and supernovae 
opens new perspectives for this scenario.

\section{Compton drag in a relativistic wind}

We consider a relativistic ($\Gamma \gg 1$) wind of plasma propagating in a 
bath of photons with typical energy $\epsilon_{\rm seed}$.
A fraction $\sim \min(1,\tau_{\rm T})$ of the photons are scattered by
the inverse Compton (IC) effect to energies $\epsilon \sim \Gamma^2 
\epsilon_{\rm seed}$, where $\tau_{\rm T}$ is the Thomson opacity of the wind.
Due to relativistic aberration, the scattered photons propagate
in a narrow cone forming an angle $1/\Gamma$ with the velocity vector 
of wind propagation.
By this process, a net amount of energy $E_{\rm CD}$ is converted from
kinetic energy of the wind to a radiation field propagating in the
direction of the wind itself, where 
$E_{\rm CD} \sim \min(1,\tau_{\rm T}) \, V \, u_{\rm rad} \,
(\Gamma^2-1)$. $V$ is volume filled by the soft photon field of
energy density $u_{\rm rad}$ swept up by the wind.

Let us assume that the GRB fireball, instead of being made by a number of 
individual shells (see e.g. Lazzati et al. 1999), is an unsteady 
(both in  velocity and density) relativistic wind, expanding from 
a central point. After an initial acceleration phase, the density 
of the outflowing wind decreases with radius as $n(r) \propto r^{-2}$, 
giving a scattering probability $\sim \min [1,(r/r_0)^{-2}]$,
where $r_0$ is the radius at which the scattering probability equals unity.
After the first scattering, the photons propagates in the same 
direction of the flow and the probability of a second scattering 
is reduced by a factor $\sim \Gamma^2$.

If such a wind flows in a radiation field with energy density
$u_{\rm rad}(r)$, the total energy transferred to the photons 
when the fireball reaches a distance $R$ is given 
by\footnote{All the calculations are made in spherical symmetry. In case
of beaming, all the quoted numbers should be considered as equivalent
isotropic values.}:
\begin{equation}
E_{\rm CD}(R) = 4 \pi \Gamma^2 \left[ \int_0^{r_0} \!\!\!\!\!\! 
u_{\rm rad}(r) r^2 dr +
\int_{r_0}^{R} \!\!\!\!\!\! r_0^2 u_{\rm rad}(r) r^2 dr \right]
\label{eq:uno}
\end{equation}
where for simplicity we assume that a constant $\Gamma$
has been reached (see also Section 3).
The transparency radius $r_0$ depends on the baryon loading of the fireball,
which is parameterized by $\eta_{\rm b} \equiv E/(M c^2)$, where
$E/M$ is the ratio between the total energy and the rest mass
of the fireball. Then $r_0$ is given 
by\footnote{Here and in the following we adopt the notation
$Q=10^xQ_x$, using cgs units}:
\begin{equation}
r_0 = 5.9 \times 10^{13} E_{52}^{1/2} \eta_{\rm \{b,2\}}^{-1/2} \, {\rm cm}.
\label{eq:due}
\end{equation}

\subsection{A simple scenario}

We initially consider a simple scenario which can illustrate
the basic features of the Compton drag effect.
Let us assume that the GRB is triggered 
at a time $\Delta t$ (of the order of a few hours)
after the explosion of a supernova (Woosley et al. 1999; Cheng \& Dai 1999).
By this time, the supernova ejecta, moving with velocity
$\beta_{\rm SN} c$, have reached a distance 
$R_{\rm SN}= v_{\rm SN} \Delta t \sim 5.4\times 10^{13} \beta_{\rm \{SN,-1\}}
(\Delta t/{\rm 5~hr})$ cm. Let us also imagine that the supernova explosion 
is asymmetric, e.g. with no ejecta in the polar directions.
Despite this asymmetry, the ejecta uniformly fill with radiation
the entire volume within $R_{\rm SN}$.
If $R_{\rm SN} < r_0$, the energy extracted by Compton drag is:
\begin{eqnarray}
E_{\rm CD} &=&  {4\pi R_{\rm SN}^3\over 3}\, \Gamma^2 \,u_{\rm rad} 
\qquad\qquad\qquad \, R_{\rm SN} \le r_0
\label{eq:tre} \\
%
%
%
E_{\rm CD} &=& {4\pi r_0^3\over 3}\, \Gamma^2 \,u_{\rm rad}
\left(3 {R_{\rm SN}\over r_0}-2 \right) \quad R_{\rm SN} > r_0.
\label{eq:qua}
\end{eqnarray}
According to Woosley et al. (1994), the average luminosity 
of a type II supernova\footnote{This luminosity decreases by a factor 
$\sim 100$ for type Ibc supernovae, while the typical frequency increases 
by a factor of 10.} during $\Delta t$ is of the order of 
$L_{\rm SN}\sim 10^{44}$ erg s $^{-1}$,
with a black body emission at a temperature $T_{\rm SN}\sim 10^5$ K.
It follows that in this case $u_{\rm rad}= aT_{\rm SN}^4\sim
7.6\times 10^5 T_{\rm \{SN,5\}}^4$ erg cm$^{-3}$ 
(consistent with $R_{\rm SN}$ assumed above).
The efficiency $\xi$ of Compton drag in extracting the fireball energy
is very large; from Eq.~\ref{eq:tre} we obtain:
\begin{equation}
\xi \, \equiv \, {E_{\rm CD}\over E}\, \sim 
0.6 \, E_{52}^{-1} \,\beta_{\rm \{SN,-1\}}^3 \left( {\Delta t \over
5~{\rm h}}\right)^3 T_{\rm \{SN,5\}}^4 \, \Gamma_2^2,
\label{eq:cin}
\end{equation}
Note here that a high efficiency can be reached even for 
$\Gamma\sim 100$.
Note that the drag itself can limit the maximum speed of the expansion
-- even in a wind with a very small barion loading -- as discussed in 
Sect.~\ref{sec:prop}.
Each seed photon is boosted by $\sim 2\Gamma^2$ in frequency, yielding
a spectrum peaking at 
$h\nu\sim 2\Gamma^2 (3kT_{\rm SN}) \sim  0.5 \Gamma_2^2 T_{\{\rm SN,5\}}$~MeV.

\subsection{A more realistic scenario}

The previous scenario requires that the GRB explodes a few hours
after a supernova. There is however a plausible alternative, 
independent of whether the massive ($>30 M_\odot$)
star (assumed to be the progenitor of the GRB) ends up with a
supernova explosion or not, and can produce a gamma--ray 
burst even if the relativistic flow and the core collapse
of the progenitor star are simultaneous or separated by 
a relatively small time interval (Woosley et al. 1999; MacFadyen, Woosley \& 
Heger 1999).

In fact there is a somewhat general consensus (e.g. MacFadyen \& Woosley 
1999; Aloy et a. 1999, but see also Khokhlov et al. 1999)
that a relativistic wind can flow in a relatively baryon free
funnel created by a bow shock following the collapse of the iron core
of the star.
Even if details of this class of models are still controversial,
the formation of the funnel seems to be a general outcome. Let us 
estimate its luminosity and more precisely the amount of energy in radiation 
crossing the funnel walls at a time $t_{\rm f}$ after its creation.
With respect to the total luminosity of the star, 
assuming it radiates at its Eddington limit $\sim L_{\rm Edd}$,
there would be a reduction by the geometrical factor equal to the ratio of the
funnel to star surfaces, which is of the order of the funnel
opening angle $\vartheta$.
However, immediately after its creation, the funnel luminosity
is much larger than $\vartheta L_{\rm Edd}$, due to two effects which
we discuss in turn.

First the walls of the funnel contain an enhanced amount of radiation
with respect to the surface layers of the star: 
the radiation once ``trapped" in the interior of the star
can escape through the funnel walls, thus enhancing the luminosity inside the
funnel for a short time.
Photons produced at a distance $s$ from the wall surface cross it
at a time $t_{\rm f}\sim \tau_{\rm s} s/c = \sigma n s^2/c$, where $\sigma$
is the relevant cross section.
This compares with the Kelvin time $t_{\rm K}\sim \sigma n R_\star^2/c$
needed for radiation to reach the star surface, yielding 
$s/R_\star \sim (t_{\rm f}/t_{\rm K})^{1/2}$.
After the time $t_{\rm f}$, the radiation produced in the layer
of width $ds$ crosses the funnel surface carrying the energy
$dE_{\rm f} \sim \vartheta \tau_\star L_{\rm Edd} ds/c$, 
the corresponding luminosity being:
\begin{equation} 
L_{\rm f}\, \sim \, {\vartheta \over 2} \, L_{\rm Edd}  \,
\left({ \tau_\star R_\star \over ct_{\rm f}} \right)^{1/2}.
\label{eq:sei}
\end{equation}
For $t_{\rm f}=100$~s and a $10M_\odot$ star
with $R_\star\sim 10^{13}$~cm ($\tau_{\star} \approx 10^8$), this 
effect can enhance the funnel luminosity by $\sim 10^4$.

Let us now consider a second plausible enhancing factor. 
If the funnel has been produced by the propagation of a bow--shock in the
star, the matter in front of the advancing front is compressed, 
with a pressure increase of ${\cal M}^2$, where ${\cal M}$ is the Mach number 
of the shock in the star. 
This (optically thick) gas then flows along the sides of
the funnel and relaxes adiabatically to the pressure of the external
matter (its original pressure). 
The result is that the funnel is surrounded by a sheath (cocoon) with density 
lower than that of the unshocked stellar material by a factor 
${\cal M}^{3/2}$ (a polytropic index of $4/3$ has 
been used in the adiabatic cooling). 
The diffusion of photons through this rarefied gas into the 
funnel is then even faster, resulting in a further 
increase of the luminosity by ${\cal M}^{3/4} \sim 200$, where a shock
speed $\beta_{\rm sf} c = 0.1c$ (MacFadyen \& Woosley 1999) and a
sound speed $\beta_{\rm s} c = 10^{-4} c$ have been assumed.
  
By taking into account both effects, the funnel luminosity
corresponds to:
\begin{equation}
L_{\rm f} \sim L_{\rm Edd}  {\vartheta \over 2}
\left( {\tau_{\star} R_\star \over c  t_{\rm f} } \right)^{1/2}
\left( {\beta_{\rm sf} \over \beta_{\rm s} } \right)^{3/4}
\sim 10^{45} \vartheta_{-1} {M_\star \over 10 M_\odot}  {\rm erg~s}^{-1},
\label{eq:set}
\end{equation}
which leads to an energy loss for Compton drag
$L_{\rm CD}  \sim  \Gamma^2 L_{\rm f}\sim 10^{49} 
\vartheta_{-1} \Gamma^2_2 (M_\star / 10 M_\odot)$,
to be compared with the observed luminosity
$\langle L_{\rm GRB}\rangle \sim 10^{49} \pi \vartheta_{-1}^2$ erg s$^{-1}$.
Here the average luminosity is considered over the 
entire burst duration:
for single pulses, we should take into account an extra factor $\Gamma^2$
in Eq.~\ref{eq:set} due to the Doppler contraction of the 
observed time.

The typical radiation temperature associated with this luminosity,
assuming a black body spectrum, is enhanced with respect to
the temperature of the star surface by 
$[L_{\rm f}/(\vartheta L_{\rm Edd})]^{1/4} \sim 
(\tau_\star R_\star)^{1/8}  (c t_{\rm f})^{-1/8}
(\beta_{\rm sf}/\beta_{\rm s})^{3/16}$.
Adopting the numerical values used above, the enhancement
is of the order of 50, corresponding to a funnel temperature 
$T_{\rm f}\sim 2\times 10^5$ K (for a surface 
temperature of the star of $\sim$5000 K).
This value is similar to the one estimated
in the simple scenario of the previous subsection and thus leads to 
similar Compton frequencies.

\section{Properties of the observed bursts}
\label{sec:prop}

If the wind is homogeneous the spectrum of the scattered photons
resembles that of the incident photons, i.e. a broad black--body
continuum peaked at a temperature $T_{\rm drag} \sim 2\, \Gamma^2 T$.
While the observed characteristic photon energy would be therefore
$\epsilon \sim 0.5 \Gamma_2^2 \, T_5 (1+z)^{-1}$~MeV, in good
agreement with the observed distribution of peak energies of BATSE
GRBs (assuming again $\Gamma=100$, see below), the spectrum
would not reproduce the observed smoothly broken power--law shape
(Band et al. 1993).  The assumptions of a perfectly homogeneous wind
and of an isothermal radiation field are however very crude, and one
might reasonably expect that different regions of the wind are
characterized by different values of $\Gamma$ and different soft field
temperatures. If we assume, e.g., that the temperature of the 
soft photon field varies with radius according to a power--law
$T(r) \propto r^{-\delta}$, the time integrated spectrum
will have a high energy power--law tail $F(\nu) \propto 
\nu^{-{3-3\delta\over\delta}}$.
In addition, the bulk Lorentz factor of the flow is likely to 
vary on a timescale much shorter than the integration time required
to obtain a spectrum with the BATSE data ($\sim 1$~s), and hence the analysed 
spectra are the superposition of drag spectra by many different Lorentz 
factors. A third effect adding power to the high energy tails of the spectrum
is the reflection of up--scattered photons in the pre--supernova
wind. This photons are scattered again by the fireball and can
reach energies of $\sim 0.5 \Gamma$~MeV $\sim 50 \Gamma_2$~MeV.
The computation of the actual spectrum resulting from all these
effects depends from many assumptions and is beyond the scope of 
this work.

The effects described above, which can increase the funnel
luminosity over the Eddington limit, take place in non--stationary conditions.
At the wind onset, it is likely that the temperature gradient 
in the walls of the funnel is large, but this is soon erased
due to the high luminosity of the walls. This causes both
the total flux and the characteristic frequency of the soft
photons to decrease, and hence a hard--to--soft trend is
expected. Moreover, it has been shown by Liang \& Kargatis (1996)
that the peak frequency of the spectrum in a single pulse at time $t$
is strongly related to the flux of the pulse integrated from
the beginning of the pulse to the time $t$. In our scenario, this 
behaviour can be easily accounted for if we consider a shell slowed down 
by the drag itself: the Lorentz factor (and hence the peak frequency of 
the spectrum) at a time $t$ is related to the energy lost by the shell, i.e.
to the integral of the flux from the beginning of the pulse to the time $t$.

The observed minimum variability time--scale is related 
to the typical size of the region containing
the dense seed photon field, which corresponds to either 
$R_\star$ or $R_{\rm SN}$ depending on which of the two scenarios 
described above applies.
The relevant light crossing time -- divided by the time compression 
factor -- is thus  
\begin{equation}
t_{\rm var} \sim {R \over c \Gamma^2} \sim 3 \times 10^{-2} \,
R_{13} \, \Gamma_2^{-2} \, {\rm s}.
\label{eq:ott}
\end{equation}
%
Longer time--scales are instead expected if
the relativistic wind is smooth and continuous.

Another interesting feature of this scenario is the possibility
that the bulk Lorentz factor of the wind is self--consistently limited
by the drag itself. 
The pressure of the
soft photons starts braking the fireball in competition
with the pressure of internal photons. The limiting Lorentz factor is 
hence reached when the internal pressure $p^\prime_{\rm fb} \propto
(T_0/\Gamma)^4$ is balanced by the pressure 
of the external photons
as observed in the fireball comoving frame $p^\prime
\propto \Gamma^2 \, T_{\rm SN}^4 \, (1+\tau_{\rm T})^{-1}$, where
$\tau_{\rm T}$ is the scattering optical depth of the wind. 
This gives:
\begin{equation}
\Gamma_{\rm lim} \sim 2\times 10^4 \, T_{\{{\rm SN, 5}\}}^{-1/2} 
\, E_{52}^{1/4} \,R_{\{0,7\}}^{-5/8} \, \eta_{\{{\rm b,5}\}}^{-1/8},
\label{eq:nov}
\end{equation}
where $R_0$ is the radius at which the fireball is released. 
Equation~\ref{eq:nov} reduces to 
$\Gamma_{\rm lim} \sim  10^4 (T_{0,11}/T_{\rm SN,5})^{2/3}$ if the 
fireball becomes transparent before reaching the coasting phase.
With such high $\Gamma$ the Compton drag would be
maximally efficient, causing the fireball to immediately decelerate 
until its $\Gamma$ reaches the value given by 
$L_{\rm CD} = L_{\rm f} \Gamma^2$, implying:
\begin{equation}
\Gamma\, = \, \left( {L_{\rm kin} \over L_{\rm f}} \right)^{1/2}
\sim \,  300 \, 
\left({L_{\{ {\rm kin},50\}} \over  L_{\{ {\rm f},45\}}}\right)^{1/2}.
\label{eq:die}
\end{equation}
These limits are in general smaller than the 
maximum $\Gamma$ set by the baryon load only, but still in   
agreement with values recently
inferred for GRB~990123 (Sari \& Piran 1999).
In addition, it is likely that the external
parts of the relativistic wind, which are in closer connection with
the funnel walls, are dragged more efficiently then the central ones,
since at the beginning the soft photons coming from the walls
can penetrate only a small fraction of the funnel before being
up--scattered by relativistic electrons. 
This may result in a polar
structured wind, with higher Lorentz factors along the symmetry axis,
gradually decreasing as the polar angle increases.

\section{Discussion}

A crucial requirement of our model is the association of GRBs with the final 
evolutionary stages of very massive stars, as these provide the 
large amount of seed photons emitted  at distances $\sim 10^13$ cm
from the central trigger, which are neeeded for the Compton drag
to be efficient.

The efficiency of conversion of bulk kinetic energy of the flow into
gamma--ray photons is large, solving the observational challenge of
gamma--ray emission being more energetic than the afterglow one
(Paczy\'nski 1999). Furthermore in this scenario
there is no requirement for either efficient acceleration in 
collisionless shocks or the presence/generation of an intense 
(equipartition) magnetic field, although Poynting flux may still 
be important in accelerating the outflow (being more efficient 
than neutrino reconversion into pairs).

We have investigated the main properties of a GRB produced by
Compton drag in a relativistic wind in a very general case.
A moderately beamed burst ($\vartheta \lesssim 10^\circ$,
Woosley et al. 1999) can be thus produced and, without any fine tuning
of the parameters, the basic features of classic GRBs are accounted for.

In particular, the peak energy of the burst emission simply reflects
the temperature of the supernova seed photons, up--scattered by the 
square of the bulk Lorentz factor.
The simplest hypothesis predicts a quasi--thermal spectrum, however it
is easy to imagine an effective multi--temperature distribution
which would depend on unconstrained quantities such as the 
variation of the spectrum of the SN photons with radius and the 
degree of inhomogeneity of the wind. 


Although in this scenario there is no requirement for internal shocks
to set up, they can of course occur, contributing a small fraction of
the observed gamma--ray flux. 
On the other hand, the wind is expected
to escape from the funnel of the star with still highly relativistic
motion, so that an external shock can be driven in the interstellar
medium and produce an afterglow, similar to the scenario 
already studied by several authors.  It is likely that 
this afterglow would develop in a non--uniform density medium, due
to the presence of the massive star wind occurring before the
supernova explosion (Chevalier \& Li 1999).

\acknowledgements 
We thank Andy Fabian, Francesco Haardt, Piero Madau and Giorgio Matt
for many stimulating discussions. DL thanks the Institute of Astronomy
for the kind hospitality during the preparation of this work. The
Cariplo Foundation (DL) and Italian MURST (AC) are acknowledged for
financial support.


\begin{thebibliography}{}

\bibitem{fl} Aloy, M. A., Mueller, E., Ibanez, J. M., Marti, J. M. \&
	MacFadyen, A., 1999, \apj~submitted (astro-ph/9911098)
\bibitem{bb} Band, D. et al., 1993, \apj, 413, 281
\bibitem{bs} Bloom, J. S., et al., 1999, \nat, 401, 453
\bibitem{so} Cheng, K. S. \& Dai, Z. G., 1999, \apj~submitted
	(astro-ph/9908248)
\bibitem{ge} Chevalier, R. A. \& Li, Z., 1999, \apj, 520, L29
\bibitem{zo} Djorgovski, S. G., Kulkarni, S. R., Bloom, J. S.,
	Goodrich, R., Frail, D. A., Piro, L., Palazzi, E., 1998,
	\apj, 508, L17 
\bibitem{dg} Eichler, D., Livio, M., Piran, T. \& Schramm, D. M., 1989,
	\nat, 340, 126
	\& Taylor, G. B., 1997, \nat, 389, 261
\bibitem{js} Fryer, C. L., Woosley, S. E., Hartmann, D. H., 1999,
	\apj~in press (astro-ph/9904122)
\bibitem{sd} Galama, T. J., et al., 1999, \apj~submitted (astro-ph/9907264)
\bibitem{ma} Ghisellini, G. \& Celotti, A., 1998, \apj, 511, L93
\bibitem{dh} Ghisellini, G., Celotti, A. \& Lazzati, D., 1999, 
	\mnras~submitted
\bibitem{z0}   Hogg, D.W., \& Fruchter, A.S., 1999, ApJ, 520, 54
\bibitem{d8} Khokhlov, A. M., H\"oflich, P., Oran, S. E., Wheeler, J. C.
	\& Wang, L., 1999, \apj~submitted (astro-ph/9904419)
\bibitem{ep} Kumar, P. \& Piran, T., 1999, \apj~submitted (astro-ph/9909014)
\bibitem{ue} Lazzati, D., Campana, S. \& Ghisellini, G., 1999a, 
	\mnras, 302, L31
\bibitem{sg} Lazzati, D., Ghisellini, D., \& Celotti, A., 1999b, 
	\mnras 309 L13
\bibitem{hc} Liang, E. P. \& Kargatis, V., 1996, \nat, 381, 49
\bibitem{k3} Liang, E. P., 1997, \apj, 491, L15
\bibitem{ee} Owens, A. et al., 1998, A\&A, 339, L37
\bibitem{hz} MacFadyen, A \& Woosley, S. E., 1999, \apj~submitted
	(astro-ph/9810274)
\bibitem{h3} MacFadyen, A, Woosley, S. E. \& Heger, A., 1999, \apj~submitted
	(astro-ph/9910034)
\bibitem{r4} M\'esz\'aros, P. \& Rees, M. J., 1997, \apj, 476, 232
\bibitem{ka} Piro, L. et al., 1999, \apj, 514, L73
\bibitem{s5} Paczy\'nski, B., 1986, \apj, 308, L43
\bibitem{sf} Paczy\'nski, B., 1998, \apj, 494, L45
\bibitem{sb} Paczy\'nski, B., 1999, to appear in: ``The Largest Explosions 
	Since the Big Bang: Supernovae and Gamma Ray Bursts" 
	(astro-ph/9909048)
\bibitem{s2} Rees, M. J. \& M\'esz\'aros, P., 1994, \apj, 430, L93
\bibitem{hf} Reichart, D. E., 1999, \apj, 512, L111
\bibitem{ss} Sari, R. \& Piran, T., 1999, \apj, 517, L109
\bibitem{p1} Sahu, K., Livio, M., Petro, L. \& Macchetto, D. F., 1997,
	IAU Circ. 6606
\bibitem{dj} Shemi,  A., 1994, \mnras, 269, 1112
\bibitem{kq} Spada, M., Panaitescu, A. \& M\'esz\'aros, P., 1999,
	\apj~submitted (astro-ph/9908097)
\bibitem{d0} Wheeler, J. C., 1999, to appear in: ``The Largest Explosions 
	Since the Big Bang: Supernovae and Gamma Ray Bursts" 
	(astro-ph/9909096)
\bibitem{ar} Woosley, S. E., 1993, \apj, 405, 273
\bibitem{he} Woosley, S. E., Eastman, R. G., Weaver, T. A. \&
	Pinto, P. A., 1994, \apj, 429, 300
\bibitem{s7} Woosley, S. E., MacFadyen, A. I. \& Heger, A., 1999,
	to appear in: ``The Largest Explosions Since the Big Bang: 
	Supernovae and Gamma Ray Bursts" (astro-ph/9909034)
\bibitem{ds} Zdziarski, A. A., Svensson, R. \& Paczy\'nski, B.,
	1991, \apj, 366, 343

\end{thebibliography}
\end{document}